\documentclass[twocolumn]{emulateapj}

\usepackage{latexsym,amssymb}
\usepackage{amsmath}
\usepackage{natbib}

\usepackage[colorlinks=true,linkcolor=blue,citecolor=blue]{hyperref}

\slugcomment{Draft \today}
\shorttitle{Did   GW170817 harbor  a pulsar?}
\shortauthors{Ramirez-Ruiz,  Andrews \&  Schr\o der}

\begin{document}

\title{Did  GW170817 harbor  a pulsar?}
\author{Enrico Ramirez-Ruiz\altaffilmark{1,2}, Jeff J. Andrews\altaffilmark{2} and Sophie L. Schr\o der\altaffilmark{2}}
\altaffiltext{1}{Department of Astronomy and
  Astrophysics, University of California, Santa Cruz, CA
  95064}
  \altaffiltext{2}{Niels Bohr Institute, University of Copenhagen, Blegdamsvej 17, 2100 Copenhagen, Denmark}
\begin{abstract}
 If the progenitor of GW170817 harbored a pulsar, then a Poynting flux dominated bow-shock cavity would have been expected to form around the traveling  binary.
The  characteristic size of this evacuated region depends strongly on the spin-down evolution of the pulsar companion,  which in turn depends on the merging timescale of the system.  If  this evacuated region is able to grow to a sufficiently large scale, then the deceleration of the  jet, and thus the onset of the afterglow, would be noticeably  delayed.  The first detection of afterglow emission, which  was uncovered 9.2 days after the $\gamma$-ray burst trigger, can thus be used to constrain the size of a pre-existing pulsar-wind  cavity. We use this information, together with a model of the  jet to place limits on the presence  of a pulsar in GW170817 and discuss the  derived constraints in the context of the observed double neutron star binary population. We find that the majority of Galactic systems that are close enough to merge within a Hubble time would have carved a discernibly large pulsar-wind cavity, inconsistent with the onset timescale of the X-ray afterglow of GW170817. Conversely, the recently detected system J1913$+$1102, which hosts a low-luminosity pulsar, provides a congruous Milky Way analog of GW170817's progenitor model. This study highlights the potential of the proposed observational test for gaining insight into the origin of double neutron star binaries, in particular if the properties of Galactic systems are representative of the overall merging population.
\end{abstract}

\keywords{binaries: close --- gamma-ray burst: general --- pulsars: general  --- stars: magnetars}
  
\section{Introduction} 
Since the discovery  of the  first double neutron star binary  by \citet{1975ApJ...195L..51H},  a total of 18 additional systems have been uncovered by radio surveys \citep{2018ApJ...854L..22S}. Although these double neutron star binaries will not merge for tens of millions of years \citep{2003Natur.426..531B,2015MNRAS.448..928K},  they make it clear that such mergers were an inescapable aftermath of binary stellar evolution \citep{2017ApJ...846..170T}.  

The  potential outcomes arising from the merger of  compact binaries was first discussed by \citet{1976ApJ...210..549L}.  They studied  the tidal disruption of a neutron star by a black hole and, although they claimed  that the  occurrence rate was  likely too rare  for them to explain the then recently  discovered  $\gamma$-ray burst sources \citep{1973ApJ...182L..85K}, enough material might be ejected from these systems to be a primary channel for heavy element nucleosynthesis.  In almost a literal sense, \citet{1976ApJ...210..549L} set the stage  for interpreting the exceptional discovery of GW170817 \citep{2017PhRvL.119p1101A}. Through the  unprecedented monitoring efforts  of GW170817 by the astronomical community \citep{2017ApJ...848L..12A,2017Sci...358.1556C}, there is now a consensus that neutron star mergers can synthesize  copious  amounts of $r$-process material \citep{2017Natur.551...80K}  and  trigger short $\gamma$-ray bursts \citep[or at least a class of them;][]{2017ApJ...848L..13A}.

Observing such systems before they merge can provide us with key information about the physics  of binary stellar evolution and  the types of gravitational-wave signals we might expect to uncover with current and future  observatories \citep{2001ApJ...556..340K}. Progress  is currently hampered by the small number of systems uncovered \citep{2018ApJ...854L..22S}, which represents  a minuscule  fraction of the total estimated population \citep{2015ApJ...800..123S}. However, the future  holds enormous promise for the study of neutron star  binaries due to the advent of large wide-field radio surveys and the nascent  field of  gravitational wave astronomy, which will allow the study of these populations in great detail. A combination
of finding more systems before and after they merge has a great potential for constraining  binary evolution but also uncovering   fundamental
physics of the merger. This motivates us  to explore potential constraints on the pre-merger properties of the progenitor system of GW170817. 

In this \emph{Letter} we examine the consequences of the hypothesis that GW170817's progenitor system hosted a pulsar. In Section \ref{sec:cav} we show that if a  pulsar is present in the system, then a Poynting flux dominated cavity would have formed in its surroundings, effectively  displacing  the interstellar medium. After a review of the relevant observations, in Section \ref{sec:aft} we examine the conditions required  for a pulsar wind cavity to grow to a sufficiently large size in order to prevent  the deceleration of the relativistic ejecta  in GW170817  and, as a result, delay  the onset of the afterglow emission \citep{2014ApJ...790L...3H}. We use this information, together with a model of the structure of the evacuated cavity  to place stringent  limits on the presence   of a pulsar in GW170817.  Finally, in Section \ref{sec:dis} we discuss the  derived constraints in the context of the observed binary neutron star population and determine the properties of the electromagnetic signatures  expected  from gravitational wave sources  harboring pulsars.

\section{The Pre-Merger Environment of GW170817}\label{sec:cav}

The optical source of  GW170817 was found by \citet{2017Sci...358.1556C} to be  associated with the early-type galaxy NGC 4993 at a distance of about 40 Mpc, with the merger site located at a
projected distance of about 2 kpc away from NGC 4993's galactic center. These observations would seem to  favor a progenitor formed $\gtrsim$ 1 Gyr ago from an isolated binary in the field, 
receiving a natal kick velocity,  $V_{\rm k}$, of about $10^2$ km s$^{-1}$ \citep{2017ApJ...850L..40A}, as expected  from merging neutron star models \citep{1998ApJ...499..520F,  2002ApJ...570..252P, 2010ApJ...725L..91K, 2011MNRAS.413..461O, 2014ApJ...792..123B}.  Based on the inferred projected distance and the properties of the host \citep{2017ApJ...850L..40A}, we expect the  typical ambient density, $n_{\rm ext}$, in which the event occurred to be around $10^{-4}$ cm$^{-3}$ \citep{2002ApJ...570..252P}, consistent with those inferred from afterglow modeling \citep[e.g.][]{2018ApJ...856L..18M}.

If the  binary progenitor of GW170817 hosted a pulsar, then a  bow shock cavity would have resulted from its interaction with the external medium, the  characteristic size of which depends primarily on the pre-merger luminosity of the system. For an  inspiraling binary with at least one pulsar, there are two widely commonly accepted mechanisms for energy dissipation. The first one is the traditional spin-down luminosity 
\begin{eqnarray}
L_{\rm p}  \approx& 1.5\times 10^{35}\; B_{9}^{-2} \tau_{{\rm p}, 9}^{-2} \text{erg s$^{-1}$},
\label{eq:lp}
\end{eqnarray}
where $\tau_{\rm p} = P/\dot P$ is the pulsar spin-down time scale ($\tau_{{\rm p},9} = \tau_{\rm p}/10^9 \text{ yr}$), $B$ is the strength of the magnetic field ($B_{9} = B/10^{9} \text{ G}$) and  we  have taken $R=12$ km  and $M = 1.4 M_{\odot}$ as fiducial parameters.  

The second one is the energy dissipation due to the torque on the binary by the magnetic field of the pulsar \citep{2012ApJ...757L...3L},
\begin{align}
L_{\rm B} &\approx 7.4\times 10^{36}\; \zeta_\phi B_{9}^2 a_{30}^{-13/2}\; \text{ erg s$^{-1}$},
\label{eq:LB}
\end{align}
where $\zeta_\phi$ is the azimuthal twist, which we take to be equal to the $\zeta_\phi \approx 1$ upper bound. This is because the flux tube will
break up when $\zeta_\phi \gtrsim 1$ and, as a result,   the linkage between the two binary 
components will be disconnected \citep{2012ApJ...757L...3L}. The orbital separation $a$ ($a_{30} = a/30 \text{ km}$), whose evolution is driven by gravitational wave emission, is given by 
\begin{align}
a_{30} \approx \left({\tau_{\rm GW} \over  1.2 \times 10^{-2}\; \text{s}}\right)^{1/4}, 
\end{align}
where $\tau_{\rm GW}=a/\dot a$ is the merging time scale.

\begin{figure}[]
\centering\includegraphics[width=0.9\linewidth,clip=true]{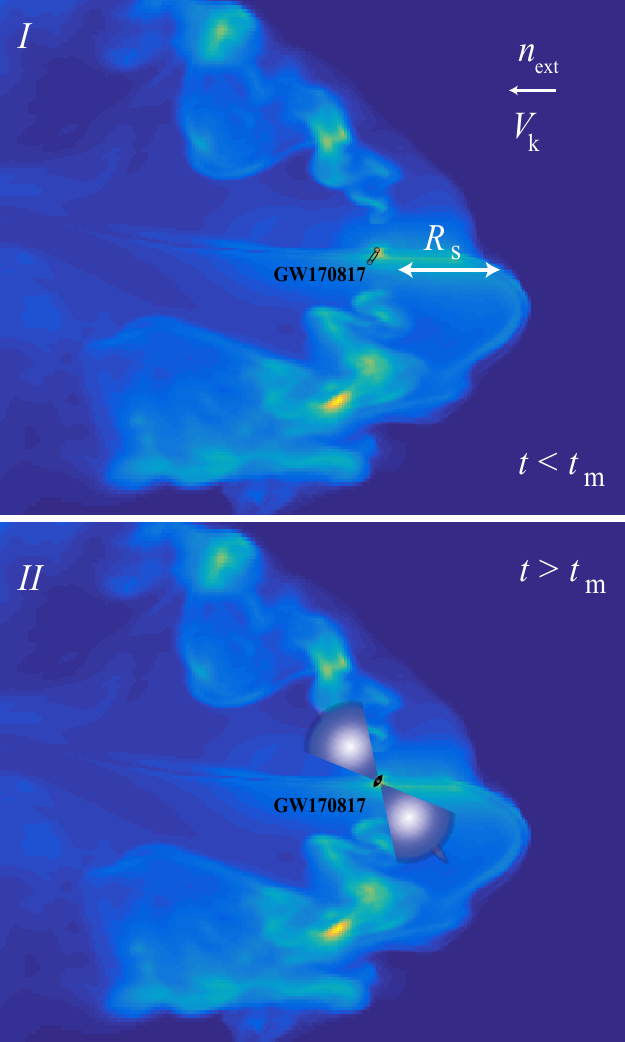}
\caption{Diagram illustrating a typical  cavity  carved by an  anisotropic and non-axisymmetric MHD pulsar wind, adapted from \cite{2018arXiv180407327B}. {\it Panel I:} $R_{\rm s}$  in the figure designates  the characteristic scale (equation~\ref{eq:rbow})   at which  the ram pressure  of the wind is balanced by that of the external medium, with a number density  $n_{\rm ext}$. The binary hosting the pulsar is assumed to be traveling through the external medium with a velocity $V_{\rm k}$ before the merger takes place (at $t=t_{\rm m}$). {\it Panel II:} Swiftly after the merger, a relativistic jet is triggered, which is expected to have a complex angular structure. The deceleration of the jet at $R>R_{\rm s}$  is assumed to power the afterglow emission in GRB 170817.}
\label{fig:diag}
\end{figure}

The binary progenitor of GW170817 would have then formed a bow shock (Fig. \ref{fig:diag}) with a  characteristic size $R_{\rm s}$ given by \citep{1996ApJ...459L..31W,2000ApJ...532..400W}
\begin{eqnarray}
R_{\rm s}  \approx  4.0 \times 10^{18}\; L_{35}^{1/2} n_{\rm ext,-4}^{-1/2} V_{{\rm k},2}^{-1} \; {\rm cm}.
\label{eq:rbow}
\end{eqnarray}
Here $L$ is the isotropic luminosity ($L_{35} = L/10^{35}{\rm erg\; s}^{-1}$), $n_{\rm ext}$ is the number density of the medium ($n_{\rm ext,-4}=n_{\rm ext}/10^{-4}\;{\rm cm}^{-3}$) and $V_{{\rm k},2} = V_{\rm k}/10^2 \text{ km s}^{-1}$. Equation \ref{eq:rbow} becomes accurate in the limit of efficient cooling and is  found to be in good  agreement with sizes of cavities carved by anisotropic and non-axisymmetric  MHD pulsar winds \citep{2007MNRAS.374..793V,2018arXiv180407327B}.  

For small kick velocities ($V_{\rm k}\lesssim$ 10 km s$^{-1}$), equation~\ref{eq:rbow}  fails to provide an accurate description for the size of the cavity and a spherically symmetric solution becomes more accurate \citep{2013ApJ...768..113M,2013MNRAS.431.2737M}. However, given that  the neutron star binary progenitor  in GW170817 likely received  a  velocity in excess of  100 km s$^{-1}$ \citep{2017ApJ...850L..40A}, we expect equation~\ref{eq:rbow} to provide us with a reasonable estimate for the characteristic size of the evacuated region.

\section{Afterglows  IN PRE-MERGER CAVITIES and Implications for GRB 170817A}\label{sec:aft}
\begin{figure}[]
\centering\includegraphics[width=0.95\linewidth,clip=false]{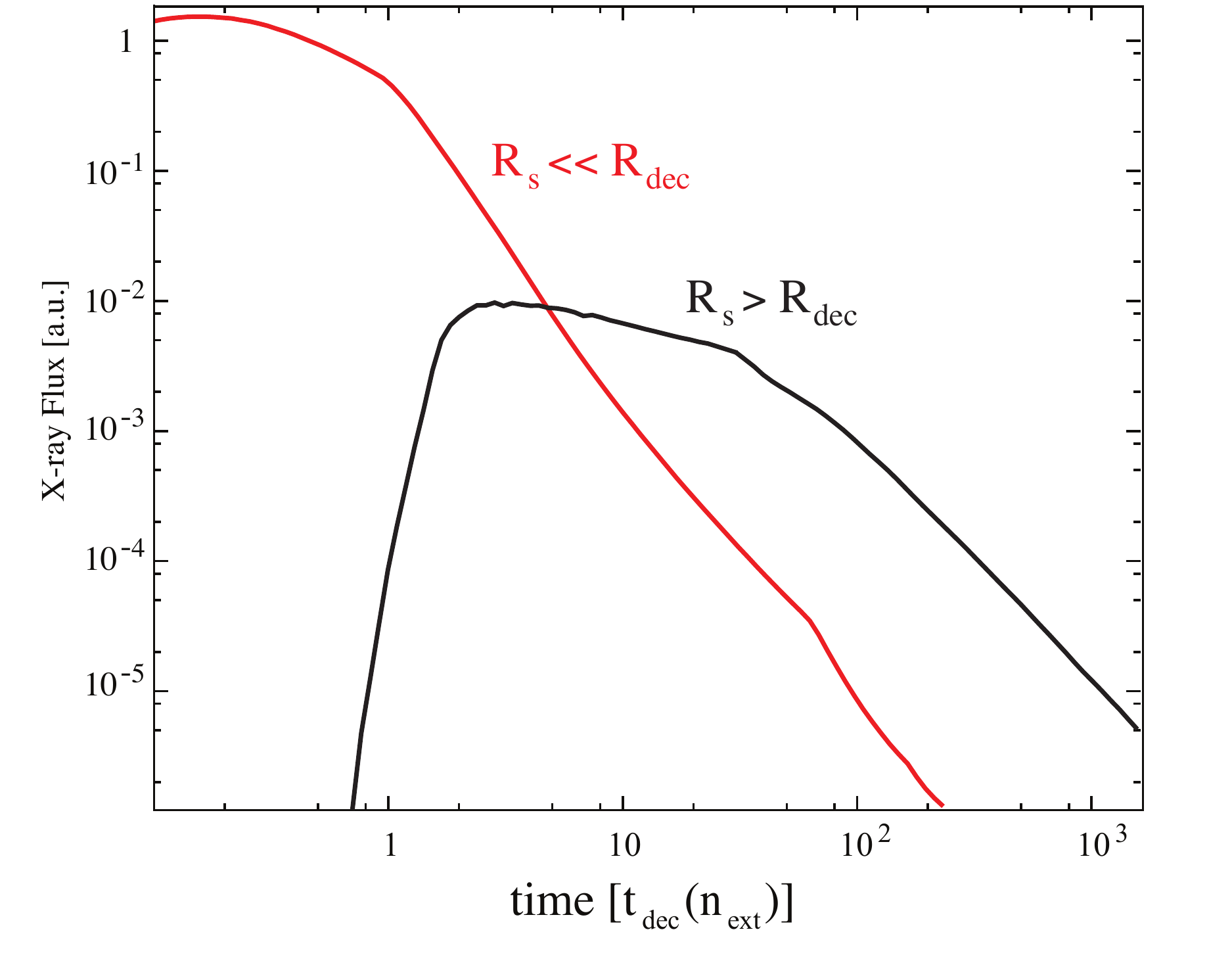}
\caption{ The X-ray afterglow lightcurves (in arbitrary units, a.u.) from sGRBs triggered by neutron star mergers. For the {\it red} curve, the X-ray emission  is computed  for a standard sGRB blastwave expanding into a  medium with constant density $n_{\rm ext}$.  The  blastwave is decelerated at $R_{\rm dec} (n_{\rm ext})$, which corresponds to $t_{\rm dec}$. The presence of a pulsar wind cavity on the X-ray afterglow lightcurve can be seen  by contrasting the {\it red} and {\it black} curves. In the pre-explosive cavity case, the blastwave expands freely within the cavity until it reaches the contact discontinuity at $R_{\rm s} = (5/2) R_{\rm dec}$. These  spherical hydrodynamic models take on the density profile of the pulsar wind cavity in the polar direction, as described in \citet{2014ApJ...790L...3H}, and provide a reasonable  description of the observed afterglow properties for relativistic material expanding along our line of sight.  Both calculations use the same microphysical parameters and assume $\Gamma=10$. A detailed description of the numerical framework, 
lightcurve calculations, and tests  are presented in \citet{2012ApJ...746..122D}. }
\label{fig:afterglow}
\end{figure}

In the study of  short $\gamma$-ray burst afterglows from neutron star  mergers one commonly considers expansion into a uniform medium \citep[e.g,][]{2002ApJ...570..252P}. In the absence of a pulsar companion,  the length scale of shock deceleration $R_{\rm dec}$  in the standard afterglow model \citep{1992MNRAS.258P..41R,1993ApJ...405..278M} is given by
\begin{align}
R_{\rm dec}  \approx  1.6 \times 10^{18}\; E_{50}^{1/3}n_{\rm ext,-4}^{-1/3}\Gamma_{5}^{-2/3}  \; {\rm cm},
\label{eq:rdec}
\end{align}
where $E_{50} = E/(10^{50} \text{erg})$ is the  isotropic-equivalent energy output of GRB 170817A and $\Gamma_{5} = \Gamma/5$ is the bulk Lorentz factor.  The values of $E$ and $\Gamma$ have been selected here to match those inferred (along our line of sight) from observations of  GW170817 \citep{2017ApJ...848L..34M, 2018PhRvL.120x1103L,2018ApJ...856L..18M,2018Natur.561..355M, 2019MNRAS.483.1247M}. This sets a
characteristic  deceleration time as measured by an  observer along our line of sight \citep{1992MNRAS.258P..41R,1993ApJ...405..278M,2006ApJ...642..354Z}
\begin{align}
t_{\rm dec} &=\bigg({1+z\over 2}\bigg ){R_{\rm dec}\over \Gamma^2c}   \nonumber \\
& \approx  27.3\; {\bigg({1+z\over 2}\bigg)} E_{50}^{1/3}n_{\rm ext,-4}^{-1/3} \Gamma_{5}^{-8/3} \; {\rm days}.
\label{eq:tdec}
\end{align} 
The relativistic expansion is then gradually slowed down, and the blastwave evolves in a self-similar manner, as shown in the {\it red} curve of Fig.~\ref{fig:afterglow}. However, since the wind of the pulsar meets the interstellar medium  at some point, the evolution of the blast wave  can be modified, as shown in the {\it black} curve of Fig.~\ref{fig:afterglow}. In this case, the shock front expands  unobstructed within the mass-evacuated pulsar wind cavity until it reaches the  contact discontinuity, which is  placed  here at $R_{\rm s} = (5/2) R_{\rm dec} (n_{\rm ext})$. The afterglow lightcurves  presented in Fig.~\ref{fig:afterglow} make use of the {\it Mezcal} special relativistic hydrodynamical calculations presented in \citet{2014ApJ...790L...3H}, in which we have assumed, for simplicity,  that  the blastwave is adiabatic and effectively spherical. 

\begin{figure}[]
\centering\includegraphics[width=0.9\linewidth,clip=false]{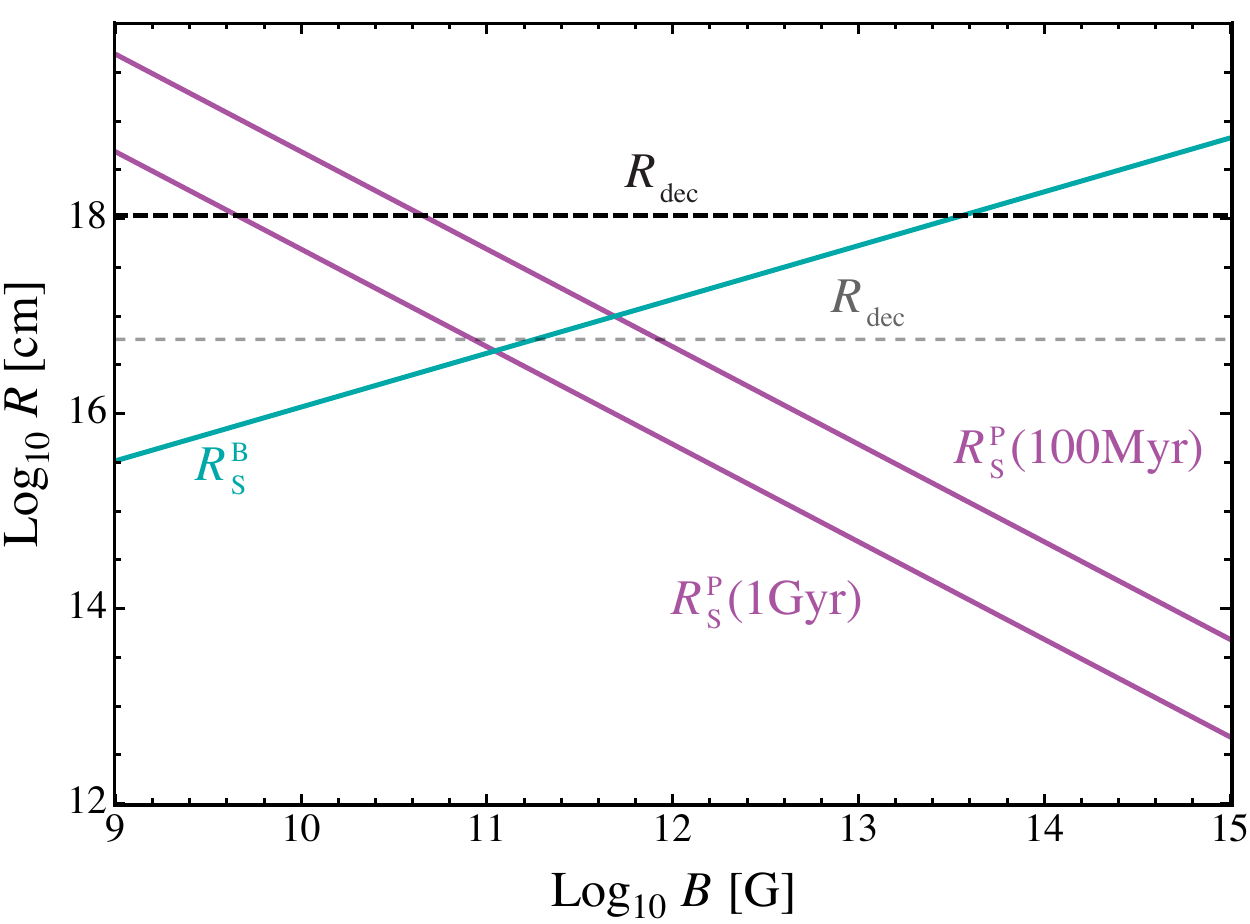}
\caption{The properties of  pre-explosive cavities in neutron  star mergers with $V_{\rm k}=100$ km s$^{-1}$ and $n_{\rm ext}=10^{-4}$ cm$^{-3}$. The relative contributions to $R_{\rm s}$ from $L_{\rm B}$ and $L_{\rm p}$ are plotted separately against the  magnetic field strength of the pulsar. The {\it purple} lines indicate the contributions from $L_{\rm p}$ for  $\tau_{\rm p} = 10^8$ yr and $10^9$ yr, roughly corresponding to the range of  expected merger delay times for GW170817 \citep{2017ApJ...850L..40A}. The {\it cyan} line shows the  expected contribution from $L_{\rm B}$ at the time of coalescence $\tau_{\rm GW} = 0$.  For comparison, we plot $R_{\rm dec} (E_{50}, \Gamma_{10})$ as a {\it black} dashed line  and $R_{\rm dec}(E_{52}, \Gamma_{300})$ as  a {\it grey} dashed line, selected  to roughly match the conditions inferred  from  the afterglow observations of  GW170817  along our line of sight and for an on-axis observer, respectively \citep{2018PhRvL.120x1103L,2018ApJ...856L..18M}.}
\label{fig:size}
\end{figure}

In general, if the  pre-merger pulsar wind  is weak enough that the blastwave would  not  be significantly slowed down by the time it expands beyond the pulsar wind cavity $R_{\rm s} \ll R_{\rm dec}$, then we  expect its  evolution as we see it to take place in a uniform medium ({\it red} curve in Fig.~\ref{fig:afterglow}). Alternatively,  we expect the presence of the pulsar wind cavity to inevitably  delay the onset of the afterglow when $R_{\rm s}\gtrsim R_{\rm dec}$  ({\it black} curve in Fig.~\ref{fig:afterglow}). It is thus useful   to examine the range of conditions that fulfill this later requirement. Taking the ratio of equations \ref{eq:rbow} and \ref{eq:rdec}, we derive
\begin{align}
\left(\frac{R_{\rm s}}{R_{\rm dec}}\right) \approx 2.5\, L_{35}^{1/2} \Gamma_{5}^{2/3} E_{50}^{-1/3} n_{\rm ext,-4}^{-1/6} V_{{\rm k},2}^{-1}.
\label{eq:ratio}
\end{align}
This ratio is  modestly  dependent on $L$ and $\Gamma$ but  weakly dependent on $n_{\rm ext}$ and $E$. \\

In Fig.\ \ref{fig:size} we compare the relative  contribution of $L_{\rm B}$ and $L_{\rm p}$ to $R_{\rm s}$ by separating it into $R_{\rm s}^{\rm B}= R_{\rm s}(L_{\rm B})$ and $R_{\rm s}^{\rm p}=R_{\rm s}(L_{\rm p})$. In order to calculate the size of the cavity produced by  $L_{\rm B}$ we must make the transformation $\tau_{\rm GW} \rightarrow \tau_{\rm GW} + R_{\rm s}^{\rm B}/c$  in order to correctly  compute  $R_{\rm s}^{\rm B}$ \citep{2014ApJ...790L...3H}.  This is because $L_{\rm B}$ propagates to the bow-shock interface within  a finite amount of time and  there is a lag of $R_{\rm s}^{\rm B}/c$ before $R_{\rm s}^{\rm B}$ can react to changes in $L_{\rm B}$. For this reason, the size of the cavity at the time of merger (equation \ref{eq:rbow}) should be computed using $L_{\rm B}$, which, after making the substitution that $a\sim(\tau_{\rm GW})^{1/4}\sim (R_{\rm s}^{\rm B}/c)^{1/4}$ in equation \ref{eq:LB}, gives $R_{\rm s}^{\rm B} \propto B^{16/29}$ (Fig.\ \ref{fig:size}). We find that $L_{\rm p}$  dominates for most realistic values of $B$ and $\tau_{\rm p}$, while  $L_{\rm B}$ becomes only relevant when  $B>10^{12}$ G (likely to be unrealistic as  it assumes no field decay). In addition, in Fig.\ \ref{fig:size} we plot two values of $R_{\rm dec}$, corresponding to the deceleration radii inferred from  afterglow observations of  GW170817 \citep{2018PhRvL.120x1103L,2018ApJ...856L..18M} along our line of sight ({\it black} dashed line) and along the  jet's axis ({\it grey} dashed line), respectively. Since $R_{\rm s}$ can in some cases exceed $R_{\rm dec}$, we expect that afterglows emanating from neutron star mergers  to noticeably  trail the prompt emission, in particular when the observer is near  the axis of the jet \citep[e.g.][]{2014ApJ...788L...8M}.  We comment further on this in Section~\ref{sec:dis}.\\

\begin{figure}
\centering\includegraphics[width=1.03\linewidth,clip=true]{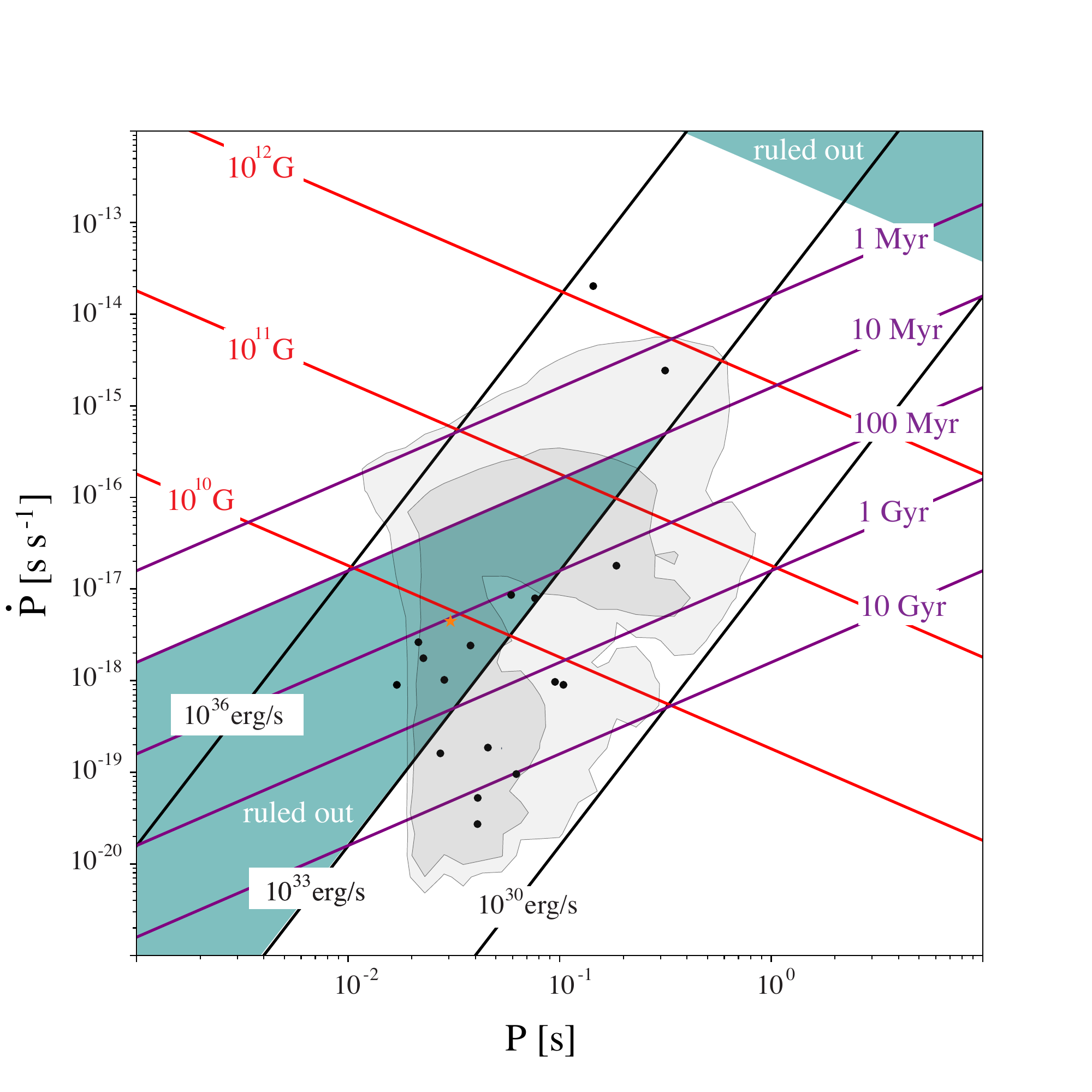}
\caption{Constraints on the presence of a pulsar in GW170817  in the $P$-$\dot{P}$ diagram. Standard lines for  $\tau_{\rm p}$ ({\it purple} lines), pulsar magnetic field $B$ ({\it red} lines), and spin-down luminosities $L_{\rm p}$ ({\it black} lines) are shown.  The {\it light blue} shaded region indicates where the condition $R_{\rm s} \gtrsim R_{\rm x}$ is satisfied. To this end,  we have adopted the  fiducial values described in equation~\ref{eq:rx} based on the afterglow solutions derived by \citet{2018PhRvL.120x1103L} and \cite{2018ApJ...856L..18M}.  For low field pulsars, we add  the  constraint that $\tau_{\rm p}  \gtrsim \tau_{\rm GW} \gtrsim 10$ Myr, based on the typical merging times of double neutron star binaries. 
The grey contours represent the  2$\sigma$ and 3$\sigma$ contours for a neutron star binary to host a pulsar companion within the given region \citep{2011MNRAS.413..461O}. Black symbols show the properties of the 17 observed pulsars in the field while the yellow star symbol shows B2127+11C,  a dynamically assembled  system located in  the globular cluster M15. }
\label{fig:ppdot}
\end{figure}

Observations summarized in \citet{2017ApJ...848L..12A}  show that the first detection of afterglow emission in GW170817 was uncovered at $t=9.2$ days at X-ray wavelengths. This can be used to set a
constraint on the  X-ray emitting  radius, $R_{\rm x}$, as measured by an  observer along our line of sight 
 \begin{align}
 R_{\rm dec} \lesssim R_{\rm x} \approx  5.5 \times 10^{17} \Gamma_{5}^{2}  \; {\rm cm}.
 \end{align}
This in turn implies that  the mass required to decelerate the ejecta was not  pushed out beyond  $R_{\rm dec}$ by the Poynting flux emanating from the  pulsar companion and, as such, it suggests that $R_{\rm s} \lesssim R_{\rm dec} \lesssim  R_{\rm x} \approx5.5 \times 10^{17} \Gamma_{5}^{2}   \; {\rm cm}$.
The fulfillment  of  the condition  $R_{\rm s}  \lesssim  R_{\rm x}$ naturally translates into a limit  on the pre-merger wind luminosity
\begin{align}
L \lesssim  2 \times 10^{33} \Gamma_{5}^{4} n_{\rm ext,-4} V_{{\rm k},2}^2\;{\rm erg\;s^{-1}}. 
\label{eq:rx}
 \end{align}
Here the bulk Lorentz factor (along our line of sight), $\Gamma$, is highly constrained by the condition that  $ 2.3\;{\rm days} \lesssim t_{\rm dec}  \lesssim 9.2\;{\rm days}$ \citep{2018ApJ...856L..18M}, which can be rewritten using equation~\ref{eq:tdec} as $1.1 E_{50}^{1/8} n_{\rm ext,-4}^{-1/8}  \lesssim \Gamma_{5}  \lesssim 2.05 E_{50}^{1/8} n_{\rm ext,-4}^{-1/8}$. This is in agreement with arguments connected with the estimated opacity of the $\gamma$-ray emitting region \citep{2019MNRAS.483.1247M},  for material moving  our line of sight.  

The non-detection of the X-ray afterglow  at $ 2.3\;{\rm days}$ can, in principle, be used to place tighter constraints on the characteristic size of the emitting region under the assumption that the afterglow data are consistent with a quasi-spherical, sub-energetic explosion (Fig~\ref{fig:afterglow}). However, given that the flow  is likely to be  collimated into jets \citep{2018PhRvL.120x1103L,2018ApJ...856L..18M}, the exact observed afterglow properties are expected  to vary depending upon the angle, $\theta_{\rm obs}$, from the symmetry axis of the jet at which they are viewed. For example, if one assumes a homogeneous sharp-edged jet \citep{2017ApJ...848L..34M}, an off-axis observer will see a rising afterglow light curve at early times (as the flow decelerates  and the Lorentz factor decreases with time) peaking when the jet Lorentz factor reaches $\sim 1/\theta_{\rm obs}$ and approaching that seen by an on-axis observer at later times. In this case, equation \ref{eq:rx} provides us with  a conservative underestimate of  the pre-merger wind luminosity. 

The constraint  on the pre-merger  luminosity should be thus  taken as an order of magnitude estimate at present and is currently consistent with the presence of an un-recycled (i.e. low luminosity) pulsar.  However, as we show below, it allows for meaningful constraints to be placed on the nature of a potential pulsar binary companion, in particular   if the properties of the Galactic neutron star binaries  are indicative of the general merging population. This is because a large  number of Galactic systems would have been expected to carve discernible cavities. If merging binaries hosting low-luminosity pulsars are, on the other hand, common, then we expect no clear delay between the onset of the afterglow and the prompt emission.  

In Fig.~\ref{fig:ppdot} we translate the condition $R_{\rm s}\lesssim R_{\rm x}$  to constraints in the $P$-$\dot{P}$ diagram.  The region of interest is separated  into two distinct zones: low field pulsars with long spin-down time scales and high field pulsars. In the latter case,  the necessary conversion $\tau_{\rm GW} \rightarrow \tau_{\rm GW} + R_{\rm s}^{\rm B}/c$ was made. For the low field pulsars, which  depend on their spin to generate pre-merger luminosity, an additional constraint should be that  $\tau_{\rm p} \gtrsim  \tau_{\rm GW}$. For this reason,  we have truncated the region of interest in Fig.~\ref{fig:ppdot}  to include low pulsar field systems with $\tau_{\rm p} \gtrsim 10$ Myr as inferred from the typical merging times of double neutron star binaries (Fig. \ref{fig:porbecc}).  For high field pulsars, on the other hand, the constraints are independent of $\tau_{\rm p}$ (equation~\ref{eq:LB}). After translating the condition that $R_{\rm s}\lesssim R_{\rm x}$ into a constraint on the pulsar $P$-$\dot{P}$ diagram, Fig.~\ref{fig:ppdot} shows that roughly half of all Milky Way double neutron star binaries fall outside the constraints. To get a sense for the likelihood that such systems are indeed representative, we have included in Fig.~\ref{fig:ppdot} the contours from the double neutron star binary population synthesis modeling of \citet{2011MNRAS.413..461O}. With the caveat that the uncertainties of population synthesis are significant, a sizable fraction of binaries lie within the  excluded region  and all of these systems are in the low field, long lived pulsar branch.

\section{Discussion}\label{sec:dis}
It is clear  from the above discussion that the uniform environment expected to surround  double neutron star binaries  at the time of merger can be altered if the system hosts a pulsar. The spin-down luminosity of the  pulsar companion is expected to be carried away in a magnetized wind that  expands into the surrounding medium, decelerating as it sweeps up ejecta from the interstellar medium, eventually  forming a bow shock cavity. The onset of the afterglow  can thus be noticeably  delayed from the prompt emission if the size of the  resultant  cavity is larger than the shock deceleration length scale in the undisturbed interstellar medium. This condition is easily satisfied for  high field pulsars $B\gtrsim 10^{13}$ G, irrespective of the spin-down age (equation~\ref{eq:LB}). However, this is only valid under the  assumption of no significant magnetic field decay. The processes regulating  pulsar field decay are still highly debated,  yet it is expected to occur on a timescale that is shorter then the lifetime of the system \citep{2011MNRAS.413..461O,2012NewA...17..594C,2014AN....335..262I,2014MNRAS.437.3863K}.  If, as expected,  magnetic field decay occurs,  then the pulsar lifetime $\tau_{\rm p}$ becomes a key parameter, as younger pulsars are  expected to produce  more extended cavities at a given magnetic field strength. 

If the magnetic field strength doesn't change significantly with time, $L_{\rm p}$ can be written in terms of the  pulsar's age (equation~\ref{eq:lp}) by assuming that its initial period $P_0$ was much shorter than the current period. From dipole radiation we know that
\begin{equation}
\dot{P} = 9.8\times10^{-19} \left( \frac{\rm B}{10^{10}\ {\rm G}} \right)^2 \left( \frac{P_0}{100\ {\rm ms}} \right)^{-1}\ {\rm s}\ {\rm s}^{-1}.
\end{equation}
Substituting this, along with $\tau_{\rm p}=P/\dot{P}$ for the spin down luminosity, into  equation~\ref{eq:rbow} we get
\begin{equation}
 R_{\rm s} = 1.6\times10^{18} n_{\rm ext, -4}^{-1/2} V_{k,2}^{-1} B_{10} P_{0, -1}^{-2} \ {\rm cm},
 \label{eq:rbow2}
\end{equation}
where $P_{0, -1} = P_0/100 {\rm ms}$. Now, from \citet{2017ApJ...846..170T} we have a relation between the (birth) spin period and the orbital period of the binary
\begin{equation}
P_0 = 36 \left( \frac{P_{\rm bin}}{{\rm days}} \right)^{0.4}\ {\rm ms},
\end{equation}
which, in turn, relates to the gravitational merger timescale 
\begin{equation}
\tau_{\rm GW} = 0.0725 \left( \frac{P_{\rm bin}}{0.1\ {\rm days}} \right)^{8/3}\ {\rm Gyr},
\label{eq:tauris}
\end{equation}
where we have assumed zero eccentricity, which provides a reasonable approximation for the merger time of most galactic double neutron star binaries. 

Under this assumption, equation~\ref{eq:rbow2}  reduces  to
\begin{equation}
R_{\rm s} = 3.5\times10^{18}\ n_{\rm ext, -4}^{-1/2} V_{\rm k,2}^{-1} B_9\ \tau_{\rm GW, Gyr}^{-0.3}\ {\rm cm}.
\end{equation}
This implies that the  eventual delay between the afterglow and the prompt emission $\approx R_s/(2c\Gamma^2)$ depends  on the merging timescale, and, as such,  offers a direct observational test on the assembly properties of the progenitor system. These constraints are likely to be more stringent for mergers in which the  jet axis is near the observer's direction and the time delay between the prompt and afterglow emission is expected to be much shorter (equation \ref{eq:tdec}).

\begin{figure}
\centering\includegraphics[width=\linewidth,clip=true]{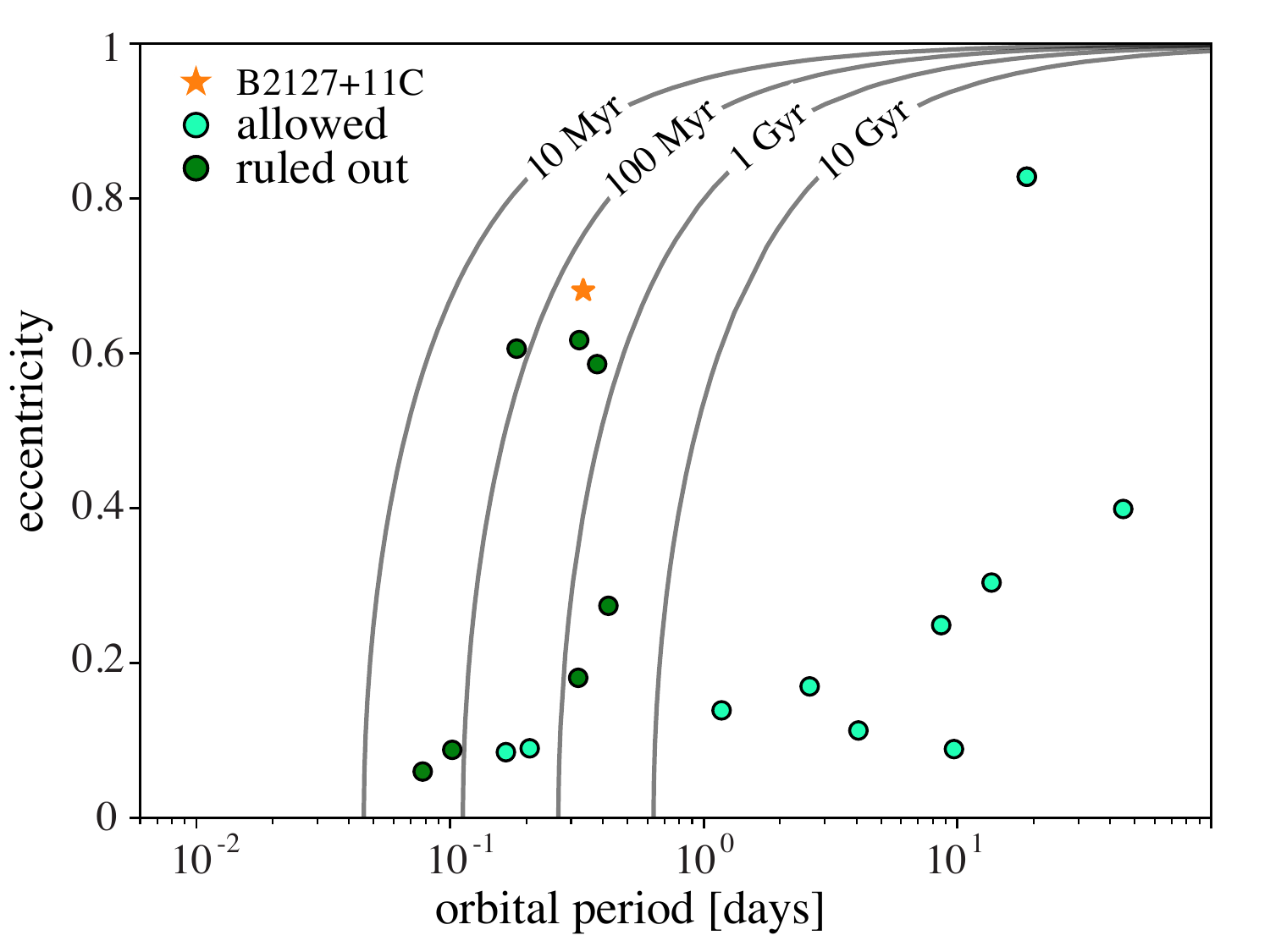}
\caption{ The orbital period and eccentricity distribution of the Milky Way DNSs, along with lines of constant merger time due to gravitational wave radiation. Excluding those systems that are too widely separated merge within a Hubble time, only two Galactic DNSs fit the $P$-$\dot{P}$ constraints defined in Fig.~\ref{fig:ppdot}. We exclude one of these, J1906$+$0746, whose nature is unconfirmed, but likely hosts a recycled pulsar that is beamed away from Earth \citep{2015ApJ...798..118V}. The other system, J1913$+$1102 \citep{2016ApJ...831..150L,2018IAUS..337..146F}, provides a potential Milky Way analog to the progenitor of GW170817 (see Section \ref{sec:dis}). }
\label{fig:porbecc}
\end{figure}

Fig. \ref{fig:porbecc} shows the distribution of orbital periods and eccentricities for the 17 double neutron star binary population in the Milky Way field, as well as B2127$+$11C, a dynamically assembled system  \citep{2006NatPh...2..116G,2010ApJ...720..953L} located in  the globular cluster M15. Lines of constant merger time due to gravitational wave radiation are indicated in black; a sizable fraction of those have orbital periods too long for the systems to merge within a Hubble time due to gravitational wave radiation; only two DNSs are allowed by the constraints on $P$ and $\dot{P}$.\footnote{This is  a relatively robust result. If, for example,  the derived  limit on the pre-merger luminosity was an order of magnitude less stringent (based on the uncertainties associated with $n_{\rm ext}$ and $V_{\rm k}$), only two additional  systems will be allowed by the  $P$ and $\dot{P}$ constraints.} One of these is the unconfirmed double neutron star binary, J1906$+$0746, that hosts a young, un-recycled pulsar. Its companion could be a massive white dwarf, but is likely a recycled pulsar that is either too weak or beamed away from Earth \citep{2015ApJ...798..118V}. The other system that satisfies the $P$-$\dot{P}$ condition, the recently detected system J1913$+$1102 \citep{2016ApJ...831..150L,2018IAUS..337..146F}, provides a potential Milky Way analog of the progenitor of GW170817. 
Given the  uncertainties in the afterglow modeling, we urge caution in the  strict application of our results. For example, if the limit  on the pre-merger wind luminosity given by  equation~\ref{eq:rx} is reduced by a factor of 6 (well within the  derived uncertainties in our estimation of the size of the cavity), the two pulsars with merging ages between  1 and 10 Gyrs in  Fig. \ref{fig:porbecc} will be able to satisfy the revised constraints in the $P$-$\dot{P}$ plane.  The precise measurement of the time
delay between  the prompt and afterglow emission can thus help provide key constraints on the pre-merger history of the binary system

\subsection{PSR J1913+1102: a Milky Way analog of the  double neutron star progenitor  of GW170817}\label{sec:analog}

Among the DNSs in the Milky Way, J1913$+$1102 is somewhat unique: whereas all other Galactic DNSs that will merge within a Hubble time have characteristic ages $<$0.5 Gyr, J1913$+$1102's characteristic age is 2.7 Gyr, a delay time of similar order to the last star formation episode of NGC 4993 \citep{2017ApJ...848L..31H,2017ApJ...849L..34P,2017ApJ...849L..16I}. Furthermore, its $P$ and $\dot{P}$ translate to $L\approx5\times10^{32}$erg s$^{-1}$ and $B\approx2\times10^9$G, characteristics that fit within the constraints we place here based on the $\approx$9 day delay between the arrival of the prompt emission and the onset of the afterglow. These constraints were derived assuming a peculiar velocity of 100 km s$^{-1}$. A dispersion measure-derived distance combined with a proper motion measurement of J1913$+$1102 indicates it is moving with a tangential velocity of $\approx$400 km s$^{-1}$; however without knowledge of the system's radial velocity it is difficult to determine the velocity of J1913$+$1102 relative to the local standard of rest. 

Despite this caveat, J1913$+$1102 produces one of the best Milky Way analogs currently available with which to compare to the progenitor of GW170817. Besides  satisfying the  total mass  requirements  \citep[as most other Galactic neutron star binaries;][]{2017PhRvL.119p1101A}, J1913$+$1102 is a large mass ratio ($q\approx1.35$) system, suggesting larger amount of tidal ejecta upon merger as estimated for GW170817 \citep{2017Sci...358.1583K,2017Natur.551...80K}. This analysis highlights the potential of the observational tool presented in this paper. Space- and ground-based observations over the coming years should allow us to uncover the detailed nature of these most remarkable binaries.

\acknowledgements
We are especially grateful to C. Holcomb and  F. De Colle for extended collaboration  and V. Kaspi and S. Ransom for discussions. We also thank the referee for many useful suggestions improving the exposition of this paper.  We are indebted  to the David and Lucile Packard Foundation, the Heising-Simons Foundation, the Danish National Research Foundation (DNRF132) and NSF (AST-1911206) for support.

% Figures ---------------


\begin{thebibliography}{}
\expandafter\ifx\csname natexlab\endcsname\relax\def\natexlab#1{#1}\fi

\bibitem[Abbott et al.(2017a)]{2017PhRvL.119p1101A} Abbott, B.~P., Abbott, R., Abbott, T.~D., et al.\ 2017, Physical Review Letters, 119, 161101 %

\bibitem[Abbott et al.(2017b)]{2017ApJ...848L..12A} Abbott, B.~P., Abbott, R., Abbott, T.~D., et al.\ 2017, \apjl, 848, L12  %

\bibitem[Abbott et al.(2017c)]{2017ApJ...848L..13A} Abbott, B.~P., Abbott, R., Abbott, T.~D., et al.\ 2017, \apjl, 848, L13 %

\bibitem[Abbott et al.(2017d)]{2017ApJ...850L..40A} Abbott, B.~P., Abbott, R., Abbott, T.~D., et al.\ 2017, \apjl, 850, L40 %

\bibitem[Barkov \& Lyutikov(2018)]{2018arXiv180407327B} Barkov, M.~V., \& Lyutikov, M.\ 2018, arXiv:1804.07327 %

\bibitem[Behroozi et al.(2014)]{2014ApJ...792..123B} Behroozi, P.~S., Ramirez-Ruiz, E., \& Fryer, C.~L.\ 2014, \apj, 792, 123 %

\bibitem[Burgay et al.(2003)]{2003Natur.426..531B} Burgay, M., D'Amico, N., Possenti, A., et al.\ 2003, \nat, 426, 531 %

\bibitem[{Chashkina \& Popov(2012)}]{2012NewA...17..594C}
Chashkina, A., \& Popov, S.~B. 2012, New Astronomy, 17, 594 %

\bibitem[Coulter et al.(2017)]{2017Sci...358.1556C} Coulter, D.~A., Foley, R.~J., Kilpatrick, C.~D., et al.\ 2017, Science, 358, 1556 %

\bibitem[De Colle et al.(2012)]{2012ApJ...746..122D} De Colle, F., Granot, J., L{\'o}pez-C{\'a}mara, D., \& Ramirez-Ruiz, E.\ 2012, \apj, 746, 122 


\bibitem[Ferdman \& PALFA Collaboration(2018)]{2018IAUS..337..146F} Ferdman, R.~D., \& PALFA Collaboration 2018, Pulsar Astrophysics the Next Fifty Years, 337, 146 %

\bibitem[{Fryer \& Kalogera(1998)}]{1998ApJ...499..520F}
Fryer, C., \& Kalogera, V. 1998, \apj, 499, 520 %

\bibitem[Grindlay et al.(2006)]{2006NatPh...2..116G} Grindlay, J., Portegies Zwart, S., \& McMillan, S.\ 2006, Nature Physics, 2, 116 %


\bibitem[Hjorth et al.(2017)]{2017ApJ...848L..31H} Hjorth, J., Levan, A.~J., Tanvir, N.~R., et al.\ 2017, \apjl, 848, L31 %

\bibitem[Holcomb et al.(2014)]{2014ApJ...790L...3H} Holcomb, C., Ramirez-Ruiz, E., De Colle, F., \& Montes, G.\ 2014, \apjl, 790, L3 %


\bibitem[Hulse \& Taylor(1975)]{1975ApJ...195L..51H} Hulse, R.~A., \& Taylor, J.~H.\ 1975, \apjl, 195, L51 %

\bibitem[{Igoshev {et~al.}(2014)Igoshev, Popov, \&
  Turolla}]{2014AN....335..262I}
Igoshev, A.~P., Popov, S.~B., \& Turolla, R. 2014, Astronomische Nachrichten,
  335, 262 %
  
\bibitem[Im et al.(2017)]{2017ApJ...849L..16I} Im, M., Yoon, Y., Lee, S.-K.~J., et al.\ 2017, \apjl, 849, L16 %
  
\bibitem[Kalogera et al.(2001)]{2001ApJ...556..340K} Kalogera, V., Narayan, R., Spergel, D.~N., et al.\ 2001, \apj, 556, 340. %

\bibitem[Kasen et al.(2017)]{2017Natur.551...80K} Kasen, D., Metzger, B., Barnes, J., Quataert, E., \& Ramirez-Ruiz, E.\ 2017, \nat, 551, 80 %


\bibitem[{Kelley {et~al.}(2010)Kelley, Ramirez-Ruiz, Zemp, Diemand, \&
  Mandel}]{2010ApJ...725L..91K}
Kelley, L.~Z., Ramirez-Ruiz, E., Zemp, M., Diemand, J., \& Mandel, I. 2010,
  \apj L, 725, L91 %

\bibitem[Kilpatrick et al.(2017)]{2017Sci...358.1583K} Kilpatrick, C.~D., Foley, R.~J., Kasen, D., et al.\ 2017, Science, 358, 1583 %


\bibitem[Kim et al.(2015)]{2015MNRAS.448..928K} Kim, C., Perera, B.~B.~P., \& McLaughlin, M.~A.\ 2015, \mnras, 448, 928 %

\bibitem[{Klus {et~al.}(2014)Klus, Ho, Coe, Corbet, \&
  Townsend}]{2014MNRAS.437.3863K}
Klus, H., Ho, W. C.~G., Coe, M.~J., Corbet, R. H.~D., \& Townsend, L.~J. 2014,
  \mnras, 437, 3863 %

\bibitem[Klebesadel et al.(1973)]{1973ApJ...182L..85K} Klebesadel, R.~W., Strong, I.~B., \& Olson, R.~A.\ 1973, \apjl, 182, L85 %

\bibitem[{Lai(2012)}]{2012ApJ...757L...3L}
Lai, D. 2012, \apj L, 757, L3 %

\bibitem[{Lattimer \& Schramm(1976)}]{1976ApJ...210..549L}
Lattimer, J.~M., \& Schramm, D.~N. 1976, \apj, 210, 549 %

\bibitem[Lazarus et al.(2016)]{2016ApJ...831..150L} Lazarus, P. and Freire, P.~C.~C. and Allen, B. et al.\ 2016, \apj, 831, 150 %

\bibitem[Lazzati et al.(2018)]{2018PhRvL.120x1103L} Lazzati, D., Perna, R., Morsony, B.~J., et al.\ 2018, Physical Review Letters, 120, 241103  %


\bibitem[Lee et al.(2010)]{2010ApJ...720..953L} Lee, W.~H., Ramirez-Ruiz, E., \& van de Ven, G.\ 2010, \apj, 720, 953 %

\bibitem[Margutti et al.(2018)]{2018ApJ...856L..18M} Margutti, R., Alexander, K.~D., Xie, X., et al.\ 2018, \apjl, 856, L18 %

\bibitem[Matsumoto et al.(2019)]{2019MNRAS.483.1247M} Matsumoto, T., Nakar, E., \& Piran, T.\ 2019, \mnras, 483, 1247 %

\bibitem[{Medvedev \& Loeb(2013{\natexlab{a}})}]{2013ApJ...768..113M}
Medvedev, M.~V., \& Loeb, A. 2013{\natexlab{a}}, \apj, 768, 113 %

\bibitem[{Medvedev \& Loeb(2013{\natexlab{b}})}]{2013MNRAS.431.2737M}
---. 2013{\natexlab{b}}, \mnras, 431, 2737 %

\bibitem[{Meszaros \& Rees(1993)}]{1993ApJ...405..278M}
Meszaros, P., \& Rees, M.~J. 1993, \apj, 405, 278 %


\bibitem[Mooley et al.(2018)]{2018Natur.561..355M} Mooley, K.~P., Deller, A.~T., Gottlieb, O., et al.\ 2018, \nat, 561, 35. %

\bibitem[Murguia-Berthier et al.(2014)]{2014ApJ...788L...8M} Murguia-Berthier, A., Montes, G., Ramirez-Ruiz, E., De Colle, F., \& Lee, W.~H.\ 2014, \apjl, 788, L8 %

\bibitem[Murguia-Berthier et al.(2017)]{2017ApJ...848L..34M} Murguia-Berthier, A., Ramirez-Ruiz, E., Kilpatrick, C.~D., et al.\ 2017, \apjl, 848, L34 %
  
\bibitem[{Os{\l}owski {et~al.}(2011)Os{\l}owski, Bulik, Gondek-Rosi{\'{n}}ska,
  \& Belczy{\'{n}}ski}]{2011MNRAS.413..461O}
Os{\l}owski, S., Bulik, T., Gondek-Rosi{\'{n}}ska, D., \& Belczy{\'{n}}ski, K.
  2011, \mnras, 413, 461%
  
  \bibitem[Palmese et al.(2017)]{2017ApJ...849L..34P} Palmese, A., Hartley, W., Tarsitano, F., et al.\ 2017, \apjl, 849, L34 %

\bibitem[{Perna \& Belczynski(2002)}]{2002ApJ...570..252P}
Perna, R., \& Belczynski, K. 2002, \apj, 570, 252 %

\bibitem[{Rees \& Meszaros(1992)}]{1992MNRAS.258P..41R}
Rees, M.~J., \& Meszaros, P. 1992, \mnras, 258, 41P %

\bibitem[Scholz et al.(2015)]{2015ApJ...800..123S} Scholz, P., Kaspi, V.~M., Lyne, A.~G., et al.\ 2015, \apj, 800, 123 %

\bibitem[Stovall et al.(2018)]{2018ApJ...854L..22S} Stovall, K., Freire, P.~C.~C., Chatterjee, S., et al.\ 2018, \apjl, 854, L22 %


\bibitem[Tauris et al.(2017)]{2017ApJ...846..170T} Tauris, T.~M., Kramer, M., Freire, P.~C.~C., et al.\ 2017, \apj, 846, 170 %

\bibitem[van Leeuwen et al.(2015)]{2015ApJ...798..118V} van Leeuwen, J., Kasian, L., Stairs, I.~H., et al.\ 2015, \apj, 798, 118 %

\bibitem[{Vigelius {et~al.}(2007)Vigelius, Melatos, Chatterjee, Gaensler, \& Ghavamian}]{2007MNRAS.374..793V}
Vigelius, M., Melatos, A., Chatterjee, S., Gaensler, B.~M., \& Ghavamian, P.
  2007, \mnras, 374, 793 %

\bibitem[{Wilkin(1996)}]{1996ApJ...459L..31W}
Wilkin, F.~P. 1996, \apj L, 459, L31 %

\bibitem[{Wilkin(2000)}]{2000ApJ...532..400W}
---. 2000, \apj, 532, 400 %

\bibitem[{Wong {et~al.}(2010)Wong, Willems, \& Kalogera}]{2010ApJ...721.1689W}
Wong, T.-W., Willems, B., \& Kalogera, V. 2010, \apj, 721, 1689

\bibitem[{Zhang {et~al.}(2006)Zhang, Fan, Dyks, Kobayashi, M{\'e}sz{\'a}ros,
  Burrows, Nousek, \& Gehrels}]{2006ApJ...642..354Z}
Zhang, B., Fan, Y.~Z., Dyks, J., {et~al.} 2006, \apj, 642, 354 %

\end{thebibliography}
\end{document}